\documentclass[twoside]{ilcws10}
\usepackage[latin1]{inputenc}
\usepackage[dvips]{graphicx,epsfig,color}
\usepackage{wrapfig,rotating}
\usepackage{amssymb,amsmath,array}

\begin{document}
\title{
Irradiation test on FD-SOI Readout ASIC of Pair-monitor} 
\author{
Yutaro Sato$^1$,
Hirokazu Ikeda$^2$,
Akiya Miyamoto$^3$,
\\
Yosuke Takubo$^1$,
Toshiaki Tauchi$^3$, 
Hitoshi Yamamoto$^1$
\vspace{.3cm}\\
1- Department of Physics, Tohoku University \\
Sendai, Miyagi 980-8578, Japan
\vspace{.1cm}\\
2- Japan Aerospace Exploration Agency (JAXA) \\
Sagamihara, Kanagawa 229-8510, Japan
\vspace{.1cm}\\
3- High Energy Accelerator Research Organization (KEK) \\
Tsukuba, Ibaraki 305-0801, Japan\\
}

\maketitle

\begin{abstract}
We fabricated a readout ASIC with the fully depleted silicon-on-insulator (FD-SOI) technology
for the pair-monitor.
The pair-monitor is a silicon pixel device
that measures the beam profile of the international linear collider.
It utilizes the directional distribution of a large number of electron-positron pairs created
by collision of bunches, and is required to tolerate radiation dose of about a few Mrad/year.
The irradiation might cause the buried oxide layer of SOI to accumulate charges
which interfere with intended functions.
We thus performed extensive irradiation tests on the prototype ASIC,
and the results are described in this report.

\end{abstract}

\section{Pair-monitor}


In the International Linear Collider (ILC), measurement of the beam size at the interaction point (IP) is important to keep high luminosity.
The nominal beam size at the IP is 639 nm ($\sigma_{x}$) $\times$ 5.7 nm ($\sigma_{y}$) $\times$ 300 $\mu$m ($\sigma_{z}$).
The goal is to achieve an accuracy of better than 10$\%$ on each dimension
which means that $\sigma_{y}$ must be measured with about 1 nm accuracy.
In our simulation study, the pair-monitor was found to attain the accuracy \cite{PMsim}.
The pair-monitor extracts information on the beam size from the directional distribution of a large number of electron-positron pairs created by collision of bunches \cite{PMorg}.
The pairs are influenced by the electromagnetic force of the oncoming beam.
Particles with charge opposite to that of the oncoming beam are focused around the oncoming beam,
while particles with the same charge as that of the oncoming beam are scattered by large angles.
The direction of the scattered particles carries information on the transverse beam shape
which determines the field pattern around the beam,
and the hit distribution of the scattered particles is measured by the pair-monitor
palced around the beam pipe at about 4 m from the IP.

We have developed a readout ASIC for the pair-monitor with the fully depleted silicon-on-insulator (FD-SOI) technology.
In this report, the irradiation test on the FD-SOI readout ASIC is described.

\section{Prototype of readout ASIC}

The pair-monitor is a single-layer disk of  silicon pixel sensor
with pixel size of about 400 $\times$ 400 $\mu$m$^{2}$ and outer radius of 10 cm.
The total number of pixels is about 200,000 per disk.
The thickness of the active region is about 200 $\mu$m
which leads to a typical signal level of about 20,000 electrons.
The estimated count rate is 2.5 MHz per one pixel at the maximum.
The beam profile may vary within a train and it is important to make measurement as a function of time within a train.
We chose to read out the pixel hit count in 16 time-slices per train.
Each count is stored on a pixel circuit with an 8-bit register.
The hit counts are then read out during the inter-train gap of 200 ms.
Since they are positioned close to the beam pipe,
a lot of low energy electron-positron pairs will deposit their energy in the pair-monitor.
The readout chip is required to tolerate radiation dose of about a few Mrad/year.

\begin{wrapfigure}{r}{0.4\columnwidth}
\centerline{\includegraphics[width=0.37\columnwidth]{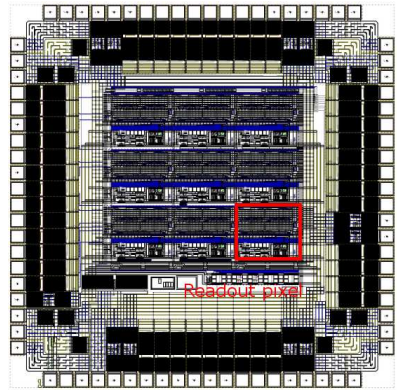}}
\caption{Layout of the readout ASIC.}\label{Fig:SOI}
\end{wrapfigure}

We developed a prototype ASIC with the OKI 0.2-$\mu$m FD-SOI CMOS process \cite{OKI} as shown in Fig. \ref{Fig:SOI}.
The readout ASIC contains nine readout pixels,
each of which consists of a pre-amplifier, shaping-amplifier, comparators, 8-bit counter and 16 count registers.
The SOI CMOS has an advantage over the bulk CMOS.
The transistors on the SOI wafer are dielectrically isolated with the field oxide and buried oxide (BOX),
and, hence, are free from possible latchup.
Since the doped impurity at the source and drain reached down to the BOX layer,
there exist no diode structure at the bottom of the source/drain implant.
It enables the SOI device to reduce the parasitic capacitance and leakage current.
In addition, since the transistor on the SOI wafer is located on a very thin silicon layer,
the energy deposit in the body by impinging radiation is relatively smaller than
that of the bulk devices;
thus the SOI device is believed to be tolerant to the single-event effect.
On the other hand, it is potentially sensitive to the total-dose effect due to the thick BOX layer ($\sim$200 nm).
The major total-dose effect is caused by radiation-induced positive charges in the BOX layer.
We can control the voltage of the handle wafer,
which is hereafter referred as the back-gate voltage.
The back-gate voltage could compensate the radiation-induced positive charges in the BOX layer.
To study the radiation tolerance, we performed irradiation test for the prototype ASIC.

\section{Irradiation test}

The readout ASIC was irradiated by the Rigaku FR-D
which mainly generates CuK$\alpha$ photons of about 8 keV.
The maximum dose was 2 Mrad.
The analog waveforms, gain, linearity, and noise level after the irradiation were compared to those before the irradiation.

\begin{figure}
\centerline{\includegraphics[width=1.0\columnwidth]{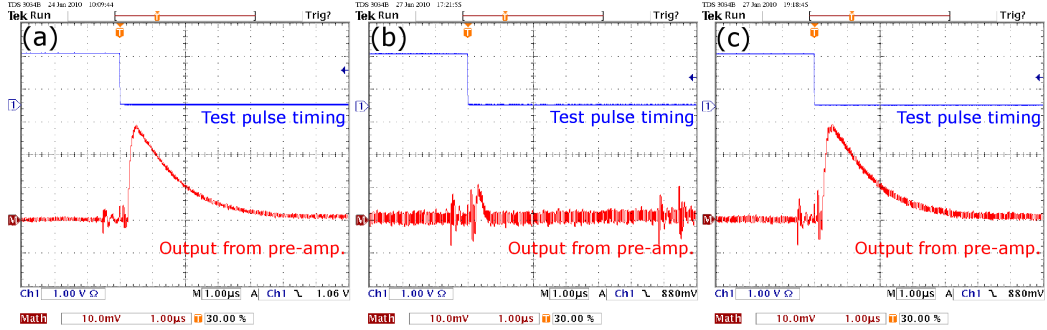}}
\caption{
Outputs from the pre-amplifier.
(a) before irradiation with fixed back gate voltage (-1.65 V).
(b) after irradiation of 300 krad with fixed back gate voltage (-1.65 V).
(c) after irradiation of 300 krad with adjusted back gate voltage (-4.72 V).
}\label{Fig:PRE}
\end{figure}

Figure \ref{Fig:PRE} shows the outputs from the pre-amplifier for the injected charges of about 20000 electrons.
The output of the pre-amplifier was gradually diminished with the irradiation,
and vanished after the irradiation of 1 Mrad.
The excess current on the power rail was observed after the 1 Mrad.
That can be understood to originate from current paths
between the source and drain of NMOS transistors,
which are induced by the positive charge accumulated on the BOX/SOI interface.
Use of a back-gate voltage can be made to compensate for the accumulated charges.
Even after the irradiation of 2 Mrad, the output of the pre-amplifier
was restored to that before the irradiation by the back-gate voltage.

\begin{wrapfigure}{r}{0.4\columnwidth}
\centerline{\includegraphics[width=0.4\columnwidth]{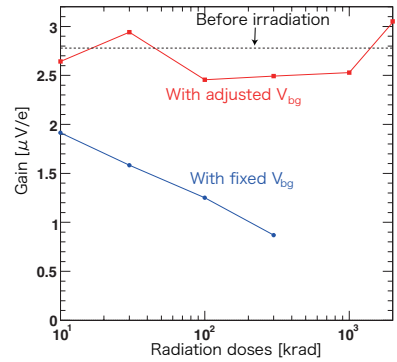}}
\caption{
Gain as a function of radiation doses.
}\label{Fig:GAIN}
\end{wrapfigure}

The gain, linearity and noise level after the irradiation were also confirmed to be restored to those
before the irradiation by adjusting the back-gate voltage.
Figure \ref{Fig:GAIN} shows the gain as a function of the radiation doses.
The gain before the irradiation was about 2.8 $\mu$V/electron.
If the back-gate voltage (V$_{\rm bg}$) is fixed to the design value of -1.65 V,
the gain was decreased with the irradiation.
The gain became below 0.9 $\mu$V after the irradiation of 300 krad.
However when the back-gate voltage was adjusted to restore the analogue waveform,
the gain after the irradiation was also restored to that before the irradiation.
The observation of the gain with the fixed back-gate voltage could not be executed
because the NMOS transistors were shorted as described above.
The fluctuation around the gain before irradiation arises from error of the adjusted back-gate voltage.

From these results we conclude that
the radiation-induced positive charges in the buried oxide can be compensated by the back-gate voltage,
and that the readout ASIC can be made to tolerate the radiation environment at the ILC.


\section{Summary}

We performed irradiation test on a readout ASIC for the pair-monitor fabricated with the OKI 0.2-$\mu$m fully depleted silicon-on-insulator (FD-SOI) process.
Even though the SOI CMOS has a number of advantages over the bulk CMOS,
if the damage to the BOX leads to the degradation of the radiation tolerance,
we cannot take the benefit from the superior properties of the FD-SOI.
We, thus, performed irradiation tests for the prototype ASIC.
We confirmed that
the radiation-induced positive charges in the buried oxide can be compensated by the back-gate voltage
and the readout ASIC survives the radiation environment of the pair-monitor at the ILC.

\section{Acknowledgments}

This study is supported in part by the Creative Scientific Research Grant No. 18GS0202 of
the Japan Society for Promotion of Science, and Dean's Grant for Exploratory Research in
Graduate School of Science of Tohoku University.



\begin{footnotesize}


\end{footnotesize}


\end{document}